\documentstyle[fullpage,12pt]{article}
\begin{document}
\baselineskip 0.65truecm
\parskip 0.5cm
\centerline{\large \sf Involutive Spacetime Distributions} 
\centerline{\sf and}
\centerline{\large \sf $p$-Brane Dynamics}
\vskip 2.5truecm
\centerline{\sf Manash Mukherjee}
\centerline{\sf Department of Physics}
\centerline{\sf Virginia Polytechnic Institute and State University}
\centerline{\sf Blacksburg, Virginia 24061 }
\vskip 3.0truecm
\centerline{\large{\sf Abstract}}
{\sf We propose 
a precise definition of multidimensional fluids generated by 
self-gravitating extended objects such as strings and 
membranes:
a $p$-dimensional perfect fluid is a smooth involutive
$p$-dimensional distribution on a spacetime, each integral manifold
of which is a timelike, connected, immersed submanifold of dimension,
$p$ -- representing the history of a $(p-1)$-dimensional extended
object. This 
geometric formulation of perfect fluids of higher dimensions 
naturally leads to 
the associated stress-energy tensor. Furthermore,
the laws of temporal evolution and symmetries of such systems are derived, in general, from the
Einstein field equations and the integrability conditions. We also present
a matter model based on a $2$-dimensional involutive distribution, and it
is shown that the stress-energy tensor for self-gravitating strings 
gives rise to a non-trivial spherically symmetric spacetime with a naked 
singularity.}
\newpage
\leftline{\large{ \sf 1 Introduction }}
\par 
The purpose of this work is to develop a general relativistic 
theory of multidimensional fluids as sources of spacetime curvature. 
The basic ingredients of such a fluid are 
($p-1$)-dimensional spatially extended objects called 
{\sf p-branes} -- where $p=1,2,3$ correspond to point particles, 
strings and membranes respectively.  More precisely, a $p$-brane is a
timelike, connected, $p$-dimensional $C^{\infty}$-manifold  
immersed in a spacetime -- representing the temporal evolution of a  
($p-1$)-dimensional extended object.   
Then, for a fixed $p$, a multidimensional fluid is defined 
by a smooth involutive $p$-dimensional distribution on a 
spacetime, each integral manifold of which is a $p$-brane.  
Thus, a multidimensional fluid 
naturally generalizes the model of a collisionless gas of point
particles by a congruence of world lines (1-dimensional distribution).   
[In this paper, all manifolds, tensor fields on them, and all maps from
one manifold to another will be $C^{\infty}$.  Also, we define a
spacetime as a non-compact, connected, oriented and time-oriented
$n$-dimensional manifold, $M$, endowed with a Lorentz metric, $g$. For
$n=4$, the corresponding spacetime will be denoted $(M^4, g)$.]  
\par
In a class of field theoretic models$^{1-4}$ of the early universe,  
$p$-branes appear as `topological defects'  
with characteristic rest-mass per unit ($p-1$)-dimensional  
spatial volume.  The existence of such extended objects could be 
a possible source 
of density perturbations, and hence may provide a causal mechanism for
generating the observed large scale structure of the universe$^5$. 
Thus any scheme, based on general relativity, for investigating the role 
of multidimensional fluids in the evolution of the universe, requires 
specification of a stress tensor, solutions of Einstein's field equations, 
description of the behaviour of other forms of matter in the vicinity of the
extended objects, and characterization of the resulting spacetimes and
their singularities.  
\par
In order to carry out this program, we need a precise form of the fluid
stress tensor, ${\sf T}$, which is formally a symmetric 
$(0, 2)$-tensor field on a spacetime $(M, g)$. Physically ${\sf T}$
replaces and unifies the concepts of energy density, momentum density,
energy flux and momentum flux. These quantities are observer dependent.
[An {\sl observer} is a future-pointing timelike curve $\gamma: I
\longrightarrow M$  ($I \subset {\bf R}$ is an open interval) such that 
$\forall s \in I$, the tangent vector $\gamma_{*s} \in T_{\gamma(s)}M$ 
satisfies $g(\gamma_{*s}, \gamma_{*s})=-1$. An {\sl instantaneous 
observer} $(x, Z)$ at $x \in M$, is a future-pointing timelike 
unit vector $Z\in T_{x}M$.]
When an
instantaneous observer $(x, Z)$ in $(M^4,g)$ measures, for instance, 
the energy density in any unit
3-volume of the local rest space
$Z^{\perp}\equiv \{X\in T_{x}M^4|g(X, Z)=0\}$, ${\sf T}(Z, Z)$
corresponds to the measured energy density. 
Also, for all known forms of matter ${\sf T}(Z, Z)\geq 0$, $\forall$
instantaneous observer $(x, Z)$ (and hence by continuity ${\sf T}(X, X)\geq 0$, 
$\forall$ causal $X\in T_x M^4$) $\forall x \in M^4$.
This operational definition uniquely specifies ${\sf T}$ in the
following sense$^6$: 
\par \noindent
\underline{\sf Theorem}:  
If the symmetric (0,2) tensor fields ${\sf T}$ and ${\sf T'}$ on a spacetime
$(M,g)$ satisfy ${\sf T}(Z,Z) = {\sf T'}(Z,Z)$ for all instantaneous  
observers $(x,Z)$ then ${\sf T} = {\sf T'}$.
\par \noindent Hence measured
energy density ${\sf T}(Z, Z)$ naturally motivates the following
\par \noindent
\underline{\sf Definition 1} : A stress-energy tensor on spacetime $M$ 
is a symmetric
$(0, 2)$-tensor field ${\sf T}$ on $M$ such that ${\sf T}(X, X)\geq 0$ for all
 causal
$X\in T_x M$, $\forall x\in M$.
\par \noindent
Based on the {\sf Theorem} above, 
we shall motivate the
definition, (1.3), of the stress tensor for a collisionless gas of
point particles (of mass $m$) in a way that is suitable for generalization to
multidimensional fluids.  
Such a fluid on $(M^4,g)$ is a
congruence of integral curves of a nowhere vanishing (energy-momentum) 
vector field, {\sf P}, on a spacetime region -- where $g{\sf
(P,P)}=-m^2$.  
This configuration of
(integral) curves is a 1-dimensional involutive distribution 
on $M^4$ (or 1-foliation of $M^4$). 
\par
The above geometric structure of a particle flow suggests that a
collection of non-colliding extended particles (non-intersecting
$p$-branes) in a spacetime, $(M^n,g)$, can be modelled 
by a $p$-dimensional ($p \geq 1$) 
foliation of $(M^n,g)$ -- where the timelike integral manifolds represent the
$p$-branes. After a brief introduction to foliations in {\sf section 2},
we then characterise a multidimensional fluid, in
{\sf section 3}, as a $p$-foliation determined locally by a {\sl nowhere
zero} decomposable $p$-form $\omega = \sigma (\tilde{V_1} \wedge \cdots
\wedge \tilde{V_p})/\mu$ where $\{{\tilde V_a}\}_{a=1}^{p}$ 
are metric dual of local vector fields $\{V_a\}$ giving
(local) bases for tangent spaces of integral manifolds of the
$p$-foliation, $\det [g(V_a,V_b)]=-\mu^2\neq 0$, 
and $\sigma$ is the characteristic rest mass per unit
$(p-1)$-dimensional spatial volume of the $p$-branes satisfying
${\cal G}(\omega,\omega)=-\sigma^2\neq 0$. Here, ${\cal G}$, is a scalar
product on the vector space of differential $p$-forms, $\Lambda^p (M^n)$ 
[see {\sf Appendix}].
\par
This local foliation $p$-form, $\omega$, together with
a {\sf density} function $\eta$ [(3.1)] allows us to
specify, locally, in a smooth
way the number of $p$-branes of the fluid
system in spacelike sections of a spacetime.
Such a description of a multidimensional fluid in terms of
$(\omega,\eta)$ naturally leads to the associated
stress tensor [(3.15)] with the following local representation:
$${\sf T}  =  \eta{\cal G}(\iota_a\omega,\iota_b\omega)e^a\otimes e^b$$
where, $\iota_a$, is the interior contraction operator on differential forms
with respect to any local basis vector fields $\{X_a\}$ with the 
corresponding dual basis $\{e^a\}$. 
\par
In {\sf section 4} we consider such a stress tensor as a possible source of
spacetime curvature, and derive 
its dynamical consequences from the Einstein
field equations. 
In particular we have shown that the
foliation $p$-form $\omega$ satisfies a `conservation law' 
[{\sf Proposition 4}] :
$$\delta (\eta\omega) = 0$$
and each integral submanifold ($p$-brane) determined by $\omega$ has
vanishing mean curvature, ${\sf H}=0$ [{\sf Proposition 3}]. 
It is also shown
that if the spacetime admits a Killing vector field ${\sf K}$ then 
the world density, $\eta$ as well as $\omega$ are 
invariant with respect to the local isometry generated by
${\sf K}$ [{\sf Proposition 1} and {\sf Proposition 2}] :
$${\sf L_K}\eta =0~~~;~~~{\sf L_K}\omega =0$$
[Here $\delta$ and ${\sf L}$ are coderivative and Lie-derivative operators
respectively].
\par
These properties can be used to solve for the foliation $p$-form,
$\omega$, and the world density, $\eta$, 
which specify the multidimensional fluid as well as the spacetime metric. 
In {\sf section 5}, we have considered a matter model 
based on a $2$-foliation where 
self-gravitating extended particles do indeed give rise to a non-trivial 
spacetime with a naked singularity.  A new
class of gravitational collapse problems is also presented.
\par \noindent
{\sf 1-Dimensional Perfect Fluids}
\par
In this subsection we motivate the definition of the stress tensor 
associated with an 1-dimensional perfect fluid 
(flow of point particles) on $(M^4,g)$, from the viewpoint of
the uniqueness {\sf Theorem} stated above. For $m\in [0,\infty)$, a
particle of mass $m$ is a future-pointing curve 
$\gamma: I\longrightarrow M^4$ such that
$g(\gamma_{*s}, \gamma_{*s})=-m^2$, $\forall s\in I$. 
Here $m\neq 0$ is the analogue of Newtonian inertial
mass and $m=0$ is allowed. The vector field, $\gamma_*$, over $\gamma$
is called
energy-momentum of the particle. Then for an instantaneous 
observer $(\gamma (s), Z)$, we
have the orthogonal decomposition of $\gamma_{*s}\in T_{\gamma (s)}M^4$ :
$$\gamma_{*s}={\bf e}Z + {\bf p}\eqno (1.1)$$
where ${\bf e}=-g(\gamma_*, Z)>0$ is the energy and ${\bf p}\in Z^{\perp}$ is 
the momentum of the particle as measured by $(x, Z)$, and hence the Newtonian
velocity of a particle with respect to $Z$ is given by
${\bf v}={\bf p}/{\bf e}\in Z^{\perp}$.
Now, if we have
enormous number of particles, each having the same mass $m\in [0, \infty)$ and
the energy-momenta, then we may describe such a system 
on $M^4$ by the following
\par \noindent
\underline{\sf Definition 2} : An 1-dimensional perfect fluid
$({\sf P},\eta,m)$ 
on $(M^4,g)$ consists of a function
$\eta : M^4 \rightarrow  [0, \infty)$ called {\sl world density} and an
energy-momentum vector field ${\sf P} : M^4 \rightarrow TM^4$ such that each
integral
curve of ${\sf P}$ is a particle of mass $m$, and the {\sf integral} of
the {\sf number density 3-form} 
$${\bf n}=\star(\eta{\widetilde {\sf P}})\eqno (1.2)$$
over a {\sf spacelike} section ${\cal D}^3\subset M^4$ defines the total
number of particles in ${\cal D}^3$. Associated with
$({\sf P},\eta,m)$ is the stress-energy tensor
$${\sf T}=\eta{\widetilde {\sf P}}\otimes{\widetilde {\sf P}}\eqno (1.3)$$
[Here $\star$ is the Hodge operator induced by the metric on $M^4$, and 
${\widetilde {\sf P}}$ is metric dual of ${\sf P}$.] 
\par \noindent
\underline{\sf Remark}: It follows from {\sf Definition 2} and the definition of
a particle (of mass $m$) 
that ${\sf P}$ is future-pointing and 
$g({\sf P}, {\sf P})=-m^2$. 
\underline{\sf Motivation for (1.3)} : {\sf T} is symmetric, smooth and
for all $X\in T_x M^4$,
${\sf T}_x(X,X)=\eta_x [g({\sf P},~ X)]^2\geq 0$. Thus {\sf T} is a
stress-energy tensor, by {\sf Definition 1}. We now explain in what
sense the measured energy density is ${\sf T}_x(Z,Z)$ for every 
instantaneous observer $(x,Z)\in T_x M^4$.
\par 
Given an observer $(x, Z)$,
(1.2), by regarding part of $T_x M^4$ (using
exponential map) 
as a part of $M^4$ for a
sufficiently small neighbourhood of $x\in M^4$,
where curvature tensor is
negligible$^6$. Then, given a set of linearly independent vectors
$X_1$, $X_2$, $X_3~ \in Z^\perp$ (rest space of $Z$), the world density
function $\eta$ can be interpreted as follows : the number of particles
measured by $(x, Z)$ in the parallelopiped 
$[X_1X_2X_3]\subset Z^\perp\subset T_x M^4$ 
is the number of integral cuves of {\sf P}
crossing the parallelopiped $[X_1X_2X_3]$, 
and is ({\sf approximately}) given by
\renewcommand {\theequation}{1.4}
\begin{eqnarray}
|{\bf n}(X_1,~X_2,~X_3)| & = & \eta_x|\Omega ({\sf P},~ X_1,~ X_2,~ X_3)|
\nonumber\\
                         & = & \eta_x {\bf e}|\Omega (Z,~ X_1,~ X_2,~ X_3)|
\end{eqnarray}
where $\Omega\equiv\star 1$ is the volume form on $M^4$,
$|\Omega (Z,~ X_1,~ X_2,~ X_3)|$ is the 3-volume of the parallelopiped
$[X_1X_2X_3]$, and from the {\sf Definition 2} and (1.1), ${\sf P}$ is given by
the following orthogonal decomposition
$${\sf P}_x ={\bf e}Z + {\bf p}\eqno (1.5)$$
An alternative way to calculate the particle number density in $Z^\perp$ is to
project, ${\bf n}_x$ [(1.2)], into $\Lambda^3 (Z^\perp)$ -- the vector
space of 3-forms on $Z^\perp$ -- by the ${\bf R}$-linear map,\\
$\Pi_Z:\Lambda^3(T_x M^4)\longrightarrow\Lambda^3(Z^\perp)$, where 
$\Pi_Z=1 + {\widetilde Z}\wedge \iota_Z$. By (1.5),
$${\bf n}_Z  \equiv  {\Pi}_Z[\star (\eta{\widetilde {\sf P}})]_x
             =  \eta_x {\bf e}\star{\widetilde Z}\eqno (1.6)$$
Then the
number of particles $(x, Z)$ measures in any unit volume of the local rest
space $Z^\perp$, is given by
$$\|{\bf n}_Z\|\equiv [{\cal G}({\bf n}_Z, {\bf n}_Z)]^{\frac 12} 
= \eta_x {\bf e}
\eqno (1.7)$$
where ${\cal G}$ [see {\sf Appendix}] is the non-degenerate symmetric
bilinear form, induced by $g$, on the
vector space $\Lambda^p (M^4)$ of differential $p$-forms. From (1.5) and (1.7)
we compute the energy density $U$ measured by $(x,Z)$ :
$$U\equiv (\eta_x {\bf e}){\bf e}\eqno (1.8)$$
Thus ${\sf T}$ is a stress tensor for the particle flow $({\sf
P},\eta,m)$, 
and by (1.3), (1.5), (1.8) we have
$${\sf T}_x(Z,Z)=\eta_x [g({\sf P},~ Z)]^2= \eta_x {\bf e}^2= U\eqno (1.9)$$
for every instantaneous observer $Z$. Hence ${\sf T}_x(Z,~Z)$ is the 
energy density of the particle flow $({\sf P},\eta,m)$ -- measured by
$(x,Z)$. Then, by the uniqueness
property,
the stress-energy tensor for a particle flow is indeed specified by (1.8).
\par \noindent
{\large{\sf 2 Involutive Distributions (or Foliations)}}
\par
In order to generalize the notion of particle flows,
we recall the following
\par \noindent
\underline{\sf Definition 3} : A $p$-dimensional smooth distribution 
${\sf D}$ on a
manifold $M^n$ is an assignment, to each point $x\in M^n$, of a $p$-dimensional
subspace ${\sf D}_x$ of $T_x M^n$.
\par \noindent {\sf Remarks}:
The smoothness of ${\sf D}$ can be expressed in two equivalent ways :
\par
(1) Every $x\in M^n$ has a neighbourhood ${\cal U}_x\subset M^n$ on which 
there exists a set of smooth
(local) vector fields $\{V_\alpha:\alpha=1,\ldots ,p\}$ 
such that the vectors $(V_\alpha)_y$ is a basis for
the subspace distinguished by the distribution ${\sf D}_y$ for every 
$y\in {\cal U}_x\subset M^n$.
Thus $\{V_\alpha\}$ are said to span the distribution, locally.
\par
(2) Every $x\in M^n$ has a neighbourhood ${\cal U}_x\subset M^n$ on which 
there exist $(n - p)$ independent smooth (local)
1-forms $\{\theta^k:k=p+1,\ldots ,n\}$ such that $\theta^k|_ 
{{\sf D}_y} = 0$ for all 
$y\in {\cal U}_x\subset M^n$. Thus $\{\theta^k\}$ are called constraint
1-forms for {\sf D}. If {\sf D} is locally spanned by $\{V_\alpha\}$,
then $\theta^k (V_\alpha)=0$.
\par
Now, an immersed submanifold ${\cal S}^p$ in $M^n$ is said to be an integral
manifold (also called a `leaf') 
of ${\sf D}$, if at each point $x\in {\cal S}^p$, its tangent space
$T_x{\cal S}^p$ coincides with the subspace ${\sf D}_x$ of $T_x M^n$. Then a
distribution ${\sf D}$ is called integrable (or involutive) if through each 
point of $M^n$ there
is an integral manifold of ${\sf D}$, and the necessary and sufficient
condition (Frobenius's integrability condition) for ${\sf D}$ to be
integrable is given by$^{8,9}$
$$[V_\alpha,~V_\beta]=f^{\gamma}_{\alpha\beta}V_\gamma\eqno (2.1)$$
for some local functions $f^{\gamma}_{\alpha\beta}$ on $M^n$. 
Or equivalently
$$d\theta^k={\lambda^k}_j\wedge \theta^j\eqno (2.2)$$
for some local 1-forms ${\lambda^k}_j$ on $M^n$. 
An integrable distribution is called a
{\sf foliation}. 
\par \noindent 
{\large{\sf 3 $p$-Dimensional Perfect Fluids}}
\par
We recall from {\sf section 1} that a $p$-brane 
is defined by a timelike, connected, $p$-dimensional manifold immersed
in a spacetime and is
distinguished by a strictly positive
parameter, $\sigma$ - the rest mass per unit 
$(p-1)$-dimensional spatial volume. Thus, we introduce the following
\par \noindent
\underline{\sf Definition 4} : 
A $p$-dimensional perfect fluid, $({\sf D},\eta,\sigma)$, in a spacetime
$(M^n,g)$  
consists of a function $\eta : M \rightarrow  [0, \infty)$,
called world density, and a smooth integrable $p$-dimensional distribution
${\sf D}$ on $M$ such that each integral manifold of
${\sf D}$ is a $p$-brane of rest mass per unit
spatial volume, $\sigma$. If the independent local vector fields
$\{V_1,\ldots,V_p\}$ span {\sf D} on the open set ${\cal U}\subset M^n$
and $\omega\equiv\sigma({\tilde V}_1\wedge\ldots\wedge{\tilde V}_p)/\mu$
where $\det [g(V_{\alpha},V_{\beta})]=-\mu^2\neq 0$, so that
${\cal G}(\omega,~\omega)=-\sigma^2$, then the {\sf integral} of the 
{\sf local number density $(n-p)$-form}
$${\bf n} = \star (\eta\omega)\eqno (3.1)$$
over a {\sf spacelike} $(n-p)$-chain ${\cal C}\subset {\cal U}$, defines the
total number of $p$-branes in ${\cal C}$. 
\par \noindent
{\sf Remarks}:
\par\noindent
(1) \underline{Locally}, ${\sf D}$ is spanned by a set of vector fields
$\{V_\alpha~:~\alpha=1,\ldots ,p\}$, which 
forms a basis for each timelike tangent space of each integral manifold
($p$-brane) of ${\sf D}$. Then $p$-branes in $({\sf D},\eta,\sigma)$ 
are {\sf locally} characterised 
by a decomposable $p$-form on $M^n$,
$\chi\equiv {\tilde V}_1\wedge\ldots\wedge{\tilde V}_p$, with 
$$\omega\equiv\sigma({\tilde V}_1\wedge\ldots\wedge{\tilde V}_p)/\mu
\eqno (3.2)$$
$$\omega^{(p)}\equiv({\tilde V}_1\wedge\ldots\wedge{\tilde V}_p)/
\mu\eqno (3.3)$$
$${\cal G}(\chi,\chi)=\det [g(V_{\alpha},V_{\beta})]=-\mu^2\eqno (3.4)$$
where $\mu$ is a strictly positive, real-valued, local function. 
The negative sign in (3.4) reflects the timelike
causal character of each integral manifold of ${\sf D}$. 
\par\noindent
(2) In (3.3), the $p$-form $\omega^{(p)}$ (hence, $\omega$ and ${\bf
n}$) is independent of the choice of the vector fields $\{V_\alpha\}$
that span {\sf D}, locally. 
Moreover, when
restricted to a $p$-brane, $\omega^{(p)}$ is the induced volume form on
the corresponding timelike integral manifold with
${\cal G}(\omega_p,~\omega_p)=-1$.
\par \noindent 
(3) A $p$-brane of rest mass per unit spatial volume, $\sigma$,
is a timelike immersion with local parametrisation
$\phi : (0, 1)^p \rightarrow M$ such that for 
$\phi (s^1,\ldots,s^p)=x\in M^n$, $\phi_{\star} (\partial_{s^\alpha})
=v_{\alpha}\in T_x M^n$ and the $p$-form 
$\omega_0\equiv\sigma({\tilde v}_1\wedge\cdots\wedge{\tilde v}_p)/\mu_0$
satisfies
$${\cal G}(\omega_0,\omega_0)=-\sigma^2\neq 0\eqno (3.5) $$
where $\det[g(v_\alpha,v_\beta)]=-\mu_0^2$, and $\sigma$ is 
the Newtonian analogue
of inertial energy per unit spatial volume of the $p$-brane$^{10}$. We also 
recall
that in a curved spacetime, a $p$-brane is self-gravitating if
$\phi$
is an extremal immersion (and hence, the mean curvature of $\phi$ vanishes).
\par\noindent
Thus given a foliation of a spacetime $(M,g)$ determined by a distribution
${\sf D}$, the corresponding multidimensioal fluid, $({\sf D},\eta,\sigma)$,
is locally characterised by the decomposable $p$-form $\omega$ on $M$. 
In order to {\sf motivate} the definition, (3.15),
of a stress-energy tensor ${\sf T}$ for such systems, we now
give an approximate local analysis to obtain the energy density 
$U_Z$ with respect to any
instantaneous observer $(x,Z)$ for all $x\in M$.
\vskip 0.5truecm
\leftline{{\sf Energy Density } $U_Z$ for ${\sf (D,\eta,\sigma)}$ :}
\par
For any $x\in M^n$, let ${\cal U}_x$ be the neighborhood of $x$
where $({\sf D},\eta,\sigma)$ is locally represented by $(\omega,\eta)$,
and consider a $p$-brane through $x$. Then, given an instantaneous
observer $(x,Z)\in T_x M^n$, $\omega$ admits the following
orthogonal decomposition with respect to $(x,Z)$:
$$\omega_x = {\tilde Z}\wedge (-\iota_Z \omega_x) 
+ \Pi_Z\omega_x\eqno (3.6)$$
where $\Pi_Z\equiv 1 + {\tilde Z}\wedge\iota_Z$ is the projection
operator, 
$\Pi_Z\omega_x$ is supported on 
$(x,Z)$'s rest-space, $Z^\perp\subset T_x M^n$ in the sense that
$[\Pi_Z\omega_x](Z)=0$, and 
${\cal G}({\tilde Z}\wedge(-\iota_Z\omega_x),\Pi_Z\omega_x)=0$. 
Now, for a $p$-brane through $x$,
its {\sf energy per unit spatial volume}, ${\cal E}_Z$, with respect to 
$(x,Z)$ is {\sf defined} by
$${\cal E}_Z=[{\cal G}(\iota_Z \omega_x, \iota_Z \omega_x)]^{\frac 12}
\eqno (3.7)$$
To motivate this definition, we look at the relevant properties of
$(p-1)$-form $\Theta\equiv\iota_Z \omega_x=(\sigma/\mu_x)\iota_Z\chi_x$.
\par\noindent
({\sf a}) From the equations (3.2) and (3.4) we have
$$\Theta = (\sigma/\mu_x)\sum_{\alpha=1}^{p} (-1)^{\alpha -1}
g(V_{\alpha x},Z){\chi}_x^\alpha\eqno (3.8)$$
where the $(p-1)$-form ${\chi}^\alpha$ is defined by
$${\chi}^\alpha\equiv {\tilde V}_1\wedge\ldots\wedge
{\tilde V}_{\alpha -1}\wedge{\tilde V}_{\alpha +1}\wedge\ldots\wedge{\tilde V_p}
\eqno (3.9)$$
Note that one of the local vector fields, $\{V_\alpha\}$,
say $V_1$, must be causal since the set $(V_\alpha)_x$ forms a basis
for a timelike subspace of $T_x M$. Then, 
$g(V_1,Z)\neq 0$ since $Z$ is timelike, and hence by (3.8)
$\Theta\neq 0$. Since, $\iota_Z\Theta = \omega_x (Z,Z)=0$ 
and $\omega_x$ is
decomposable,
the dimension of the characteristic subspace$^8$ of the non-zero $(p-1)$-form
$\Theta$ is
$(n-p+1)$, and hence $\Theta$ is also decomposable and can be written as
$$\Theta = (\sigma/\mu_x)({\tilde Y}_1\wedge\ldots\wedge{\tilde Y}_{p-1})
\eqno (3.10)$$
where $\{Y_1,\ldots,Y_{p-1}\}$ are linearly independent vectors in
$T_x M$.
Now, $\iota_Z\Theta =0$ and the linear independence of the $Y_a$'s imply
$$g(Y_a,Z)=0~~\forall a=1,\ldots,p-1\eqno (3.11)$$
From (3.11) it follows $Y_a\in Z^\perp$, and hence each $Y_a$ is spacelike and
${\cal G}(\Theta,\Theta)~>~0$.
\par\noindent
({\sf b}) For any normal field $N$ where $g(N,V_\alpha)=0$
for all $\alpha$, we find $\iota_N\Theta = \iota_N\iota_Z\omega =0$. Then by
$(3.10)$ it follows that $g(Y_a,N)=0$ and hence each spacelike $Y_a$ also
belongs to ${\cal V}_x^p$ - the tangent space of a $p$-brane through
$x\in M$, which is
spanned by the set of independent vectors $\{V_{1x},\ldots,V_{px}\}$.
\par\noindent
({\sf c}) From the above characterisation of $\Theta$ by (3.8), (3.10) and (3.11)
it is now clear that the non-zero $(p-1)$-form $\Theta$
is constructed from a set
of linearly independent $(p-1)$ {\sl spacelike}
vectors, $\{Y_a\in Z^\perp~:~a=1,\ldots,p-1\}$, which also belong to the
(Lorentzian) tangent space ${\cal V}_x^p$ of a 
(timelike) $p$-brane through $x\in M$. Thus in $Z^\perp\subset T_x M^n$, 
the $(p-1)$-plane formed by
$\{Y_a\}$ [and hence
$\iota_Z\omega_x =
(\sigma/\mu_x)\iota_Z\chi_x$] 
represents the `spatial extension' of a
$p$-brane through $x\in M$ with respect to an observer $(x,Z)$ and 
it follows that
$$A_Z\equiv [{\cal G}(\iota_Z\chi_x,\iota_Z\chi_x)]^{\frac 12}>0$$
is the spatial volume of the $p$-brane in $Z^\perp$. 
Now, projecting the foliating
$p$-form $\chi$ onto $Z^\perp$ [as in (3.6)] and using
${\cal G}(\chi,\chi)=-\mu^2$ [(3.4)] we also have
\newline
$-(\mu_x)^2 = - (A_Z)^2 + (Q_Z)^2$ - where
$$Q_Z\equiv [{\cal G}(\Pi_Z\chi_x,\Pi_Z\chi_x)]^{\frac 12}\geq 0$$
is the volume of the $p$-plane spanned by $\{V_\alpha\}$, 
when projected into
$Z^\perp$. From the above relation connecting $\mu_x$, $A_Z$ and $Q_Z$ we may
define
$$\gamma(Z)\equiv (A_Z/\mu_x)
           = [1 - (Q_Z/A_Z)^2]^{-{\frac 12}}$$
and from (3.10) compute
\begin{eqnarray*}
[{\cal G}(\Theta,\Theta)]^{\frac 12} & = & (\sigma/\mu_x)A_Z\\
                                     & = & \sigma\gamma (Z)\\
& = & \sigma [1 - (Q_Z/A_Z)^2]^{-{\frac 12}}
\end{eqnarray*}
If an observer $Z$ belongs to the tangent space 
${\cal V}_x^p$
of a $p$-brane, 
then $\Pi_Z\omega_x=0$, and hence $Q_Z=0$ and $\gamma (Z)=1$. In this case we have
$$[{\cal G}(\Theta,\Theta)]^{\frac 12}\bigg |_{Z\in {\cal V}_x^p} = \sigma$$
where $\Theta = \iota_Z \omega_x$. Thus
${\cal E}_Z\equiv [{\cal G}(\iota_Z \omega_x,\iota_Z \omega_x)]^{\frac 12}$ is
indeed the energy per unit spatial volume of a $p$-brane  with respect to
any instantaneous observer $(x,Z)$.
\par
Now, given an instantaneous observer $(x,Z)$, we can find a `sufficiently
small' neighborhood of $x\in M$ [where curvature tensor is negligible]
which (by exponential map) can be regarded$^6$ as part of $T_x M^n$. 
In such a neighborhood of $x$,
we compute the `number density' with respect to 
$(x,Z)$ by projecting ${\bf n}_x$ [(3.1)] into $\Lambda^{n-p}(Z^\perp)$
[as in (1.5)]:
$${\bf n}_Z \equiv  {\Pi}_Z[\star (\eta\omega)]_x
           = \eta_x \star [{\tilde Z}\wedge (-\iota_Z \omega_x)]\eqno
(3.12)$$
where we used the identity ${\tilde Z}\wedge\star\Psi
=(-1)^{k-1}\star[\iota_Z\Psi]$ for $\Psi\in\Lambda^{k}(T_x M^n)$.
Then, with respect to $(x,Z)$, the number of $p$-branes of ${\sf D}$
intercepted by unit volume of an $(n-p)$-plane in
$T_x M^n$ -- orthogonal to the $p$-plane represented by the nonzero
$p$-form ${\tilde Z}\wedge(-\iota_Z\omega_x)$ -- is approximately
given by [as in (1.6)]
$$\|{\bf n}_Z\|\equiv [{\cal G}({\bf n}_Z, {\bf n}_Z)]^{\frac 12}
=\eta_x [{\cal G}(\iota_Z \omega_x, \iota_Z \omega_x)]^{\frac 12}
\eqno (3.13)$$
\par\noindent
Finally, taking the product of $\|{\bf n}_Z\|$ in (3.13) and 
${\cal E}_Z$ [(3.7)]
we find the energy density $U_Z$
(that $(x,Z)$ measures) of the
fluid $({\sf D},\eta,\sigma)$ locally characterised by $\omega$ :
$$U_Z = \eta_x{\cal G}(\iota_Z \omega_x,\iota_Z \omega_x)\geq 0\eqno (3.14)$$
\vskip 0.5truecm
\leftline{{\sf Stress Tensor} for ${\sf (D,\eta,\sigma)}$ :}
\par
Now $U_Z$ is supposed to be equal to ${\sf T}(Z,Z)$ for every observer $Z$ (see
 our
discussion before {\sf Definition 1}) for any given form of the stress
tensor ${\sf T}$. Then the structure of $U_Z$ in (3.14) suggests the following
definition of the {\sl stress tensor} ${\sf T}$ for a multidimensional
fluid 
$({\sf D},\eta,\sigma)$ with the local representation  
$${\sf T} = \eta{\cal G}(\iota_a\omega,\iota_b\omega)e^a\otimes e^b\eqno
 (3.15)$$
where $\{e^a\}$ are the local basis 1-forms (on $M$) dual to $\{X_b\}$ such that
$e^a(X_b)={\delta^a}_b$ for $a,~b~=1,\ldots,n$ and $\iota_a\equiv\iota_{X_a}$.
It is clear that ${\sf T}$ is symmetric, and for any observer $(x,Z)$
\renewcommand {\theequation}{3.16}
\begin{eqnarray}
{\sf T}_x (Z,Z) & = & \eta_x {\cal G}(\iota_a\omega_x,\iota_b\omega_x)e^a (Z)e^b
 (Z)\nonumber\\
       & = & \eta_x {\cal G}(\iota_Z\omega_x,\iota_Z\omega_x)\geq 0
\end{eqnarray}
Then by continuity ${\sf T}_x (W,W)\geq 0$ for all causal $W\in T_x M$ and hence
${\sf T}$ is
a stress tensor. Furthermore, the equations (3.14) and (3.16) show that the
energy density $U_Z$ is equal to ${\sf T}_x(Z,Z)$ for every instantaneous
observer $(x,Z)$. 
Hence by the uniqueness
property, the stress tensor for the fluid, $({\sf D},\eta,\sigma)$, is specified by the
equation (3.15). We also remark that replacing $\omega$ in (3.15) by
the energy-momentum 1-form, ${\tilde P}$, reproduces the stress tensor for the
particle flows.
\par
The stress tensor ${\sf T}$ defined in (3.15) for the fluid 
$({\sf D},\eta,\sigma)$ can be written in a form which is more 
suggestive as well as
convenient for applications. From (3.4) we have
\renewcommand {\theequation}{3.17}
\begin{eqnarray}
\det[g(V_\alpha,V_\beta)] & \equiv & {\cal G}({\tilde V}_1\wedge\ldots\wedge
{\tilde V}_p,~{\tilde V}_1\wedge\ldots\wedge{\tilde V}_p)\nonumber\\
& \equiv & {\cal G}(\chi,\chi)\equiv -\mu^2\neq 0
\end{eqnarray}
Now, expanding $\iota_a\omega$ in (3.15) in terms of $\chi^\alpha$
[(3.9)],
$$\iota_a\omega= (\sigma/\mu)\sum_{\alpha=1}^{p} (-1)^{\alpha -1}
g(V_\alpha,X_a){\chi}^\alpha\eqno (3.18)$$
and inserting (3.18) in (3.15) we have
\renewcommand {\theequation}{3.19}
\begin{eqnarray}
{\sf T} & = & \eta (\sigma/\mu)^2 \sum_{\alpha,\beta}(-1)^{\alpha +\beta}
{\cal G}(\chi^\alpha,\chi^\beta)
g(V_\alpha,X_a)g(V_\beta,X_b)e^a\otimes e^b\nonumber\\
& = & \eta (\sigma/\mu)^2 \sum_{\alpha,\beta} C^{\alpha\beta}
{\tilde V_\alpha}\otimes{\tilde V_\beta}
\end{eqnarray}
where
$C^{\alpha\beta}\equiv (-1)^{\alpha+\beta}{\cal G}(\chi^\alpha,\chi^\beta)$
is the cofactor of the matrix element
${\hat g}_{\alpha\beta}\equiv [g(V_\alpha,V_\beta)]$
and from (3.17), the inverse of ${\hat g}_{\alpha\beta}$ is given by
${\hat g}^{\alpha\beta}\equiv C^{\alpha\beta}/(-\mu^2)$.
From these definitions and (3.19), we have
$${\sf T}=-(\sigma^2 \eta) {\hat g}^{\alpha\beta}{\tilde V_\alpha}\otimes{\tilde
 V_\beta}
\eqno (3.20)$$
Defining
$${\hat g}\equiv{\hat g}^{\alpha\beta}{\tilde V_\alpha}\otimes{\tilde V_\beta}
\eqno (3.21)$$
we note that ${\hat g}$ is simply a rank-2 symmetric tensor field on the
 Lorentzian
manifold $(M^n,g)$ and constructed only from the foliating vector fields
$\{V_\alpha\}$. Now computing the components of ${\hat g}$ on a leaf $L_p$
(whose tangent space is spanned by $\{V_\alpha\}$), we find from (3.21)
$${\hat g}(V_\lambda,V_\nu) = {\hat g}_{\lambda\nu}\eqno (3.22)$$
where we used the fact [see the definitions below (3.19)] that
${\hat g}^{\alpha\beta}$ is the inverse of
the matrix element ${\hat g}_{\alpha\beta}\equiv g(V_\alpha,V_\beta)$.
${\hat g}^{\alpha\beta}$ exists since by (3.17)
${\hat g}$ is non-degenerate :
$$\det[g(V_\alpha,V_\beta)]\equiv\det[{\hat g}_{\alpha\beta}]=-\mu^2\neq 0
\eqno (3.23)$$
Then from (3.22) and (3.23), restriction of
the symmetric tensor field ${\hat g}$ [(3.21)] onto each leaf $L_p$ defines
a metric on $L_p$ - induced by the Lorentzian metric $g$ on $M$.
Since the leaves of the foliation are connected and timelike,
${\hat g}$ - restricted
to a leaf - is also a Lorentzian metric of constant index.
Thus we have a  simple interpretation of (3.20) that ${\sf T}$ is proportional
 to
a metric ${\hat g}$ on each integral submanifolds, and from (3.20)-(3.21)
$${\sf T}=-\rho {\hat g}\eqno (3.24)$$
where the {\sl positive} function (on $M$),
$\rho\equiv {\sigma^2} {\eta}$,
is the energy density measured by all observers tangential to the leaves $L_p$.
\par
As we mentioned earlier [see remark(2) below {\sf Definition 3} in
{\sf section 2}], a $p$-foliation of an $n$-dimensional manifold may also be 
prescribed by $(n-p)\equiv q$ constraint 1-forms $\{\theta^i\}$ where
$\theta^i (V_\alpha)=0$ and by
suitable linear combinations from the linearly independent set
$\{{\tilde \theta^i}\}$ we can get an orthonormal set of $q$ normal fields
$\{N_k\}$ such that
\begin{eqnarray*}
g(N_i,N_j) & = & \delta_{ij}\\
g(N_i,V_\alpha) & = & 0
\end{eqnarray*}
Then
the sets $\{V_\alpha\}$ and $\{N_k\}$ together form a local basis for the 
tangent spaces of
$M$ and $\{{\tilde N_k}\}$ are the new constraint 1-forms satisfying the
integrability condition (2.2).
In terms of these normal fields (3.24) can be written as
$${\sf T}=-{\rho} {(g-\sum_{k=1}^{q}
{\tilde N_k}\otimes{\tilde N_k})}~~~;~~~q\equiv (n-p)\eqno (3.25)$$
\par \noindent 
\leftline{\large{\sf 4 Dynamics and Symmetries of Multidimensional
Fluids}}
\par
The stress tensor, ${\sf T}$ [in equivalent forms (3.15), (3.24), (3.25)], 
associated with a $p$-dimensional fluid is simply a symmetric tensor 
field on $(M^n,g)$.  However, if the spacetime admits Killing vector 
fields, {\sf K}, then {\sf T} may acquire new symmetries through the
Einstein field equation   
$${\sf G=T}\eqno (4.1)$$
where ${\sf G\equiv Ric} - {\frac 12}g{\sf R}$ is the Einstein tensor of the
spacetime $(M^n,g)$, constructed from the Ricci tensor ${\sf Ric}$ and the scalar
curvature ${\sf R}$.  Since Killing vector fields generate local isometries 
of $(M,g)$, we have$^7$ ${\sf L_K Ric = 0} = {\sf L_K R}$
where ${\sf L_K}$ is the Lie derivation with respect to ${\sf K}$. Then from the
definition of the Einstein tensor ${\sf G}$ it also follows ${\sf L_K G = 0}$,
and hence by (4.1), ${\sf L_K T = 0}$.  
\par \noindent 
\underline{\sf Proposition 1}: If ${\sf K}$ is a Killing vector field
and ${\sf (D,\eta,\sigma)}$ is a $p$-dimensional fluid on $(M^n,g)$, then
the world density function $\eta$ satisfies ${\sf L_K}\eta = 0$.
\par\noindent
\underline{\sf Proof}: 
Taking {\sf trace} of both sides of (4.1) with {\sf T} given by (3.25),
we find 
$$\rho = ({\frac{1}{p}})({\frac{n}{2}} - 1){\sf R}\eqno (4.2)$$
where $n$ and $p$ are the dimensions of $M$ and the distribution 
${\sf D}$, respectively. Since ${\sf L_K R} = 0$, and
$\rho\equiv\sigma^2\eta$, it follows from (4.2) that ${\sf L_K}\eta
=0$.  $\Box$ 
\par\noindent
\underline{\sf Corollary 1}:
If ${\sf T}$ in (4.1) is the fluid stress tensor
${\sf T}=-\rho {\hat g}$ [(3.24)], and ${\sf K}$ is a Killing vector
field on $(M,g)$, then ${\sf L_K}{\hat g} =0$.
\par\noindent
\underline{\sf Proof}: Since ${\sf L_K T}=0$, 
it follows that
$-({\sf L_K} \rho){\hat g} - \rho({\sf L_K} {\hat g}) = 0$. Then the
corollary follows from the fact that  
${\sf L_K \rho = 0}$, by {\sf Proposition 1}, and $\rho \neq 0$.
$\Box$ 
\par 
We now discuss the significance of {\sf Corollary 1} which
suggests that the symmetries of a spacetime $(M^n,g)$ are also the
symmetries of the leaves $\{L_p\}$ of a given foliation of $M$.
First, we prove a consequence
of ${\sf L_K} \hat{g} = 0$, where  
${\hat g}  =  g-\sum_{k=1}^{q}{\tilde N_k}\otimes{\tilde N_k}$ [(3.25)],
the set $\{N_k\}$ is normal to the foliating vector fields 
$\{V_\alpha\}_{\alpha =1}^{p}$,
$g(N_i,N_j)=\delta_{ij}$ and $q\equiv (n-p)$. 
\par \noindent
\underline{\sf Lemma 1}: If ${\sf L_K}{\hat g}=0$ where {\sf K} is 
Killing, then
$\displaystyle{{\sf L_K} N_j =\sum_{k\neq j} a^k N_k}$
where the $a^k$'s are real numbers. 
\par\noindent
\underline{Proof}: Taking the Lie derivative of $\hat{g}$ we have 
${\sf L_K} (g-\sum_{i=1}^{q}{\tilde N_i}\otimes{\tilde N_i})=0$.  
Since ${\sf K}$ is Killing (and hence ${\sf L_K} g=0)$,
$$\sum_{i=1}^{q}({\sf L_K}{\tilde N_i}\otimes{\tilde N_i}
+ {\tilde N_i}\otimes{\sf L_K}{\tilde N_i})=0\eqno (4.3)$$
Now, evaluating the symmetric tensor in (4.3) on $\{V_\alpha,N_j\}$
$$ 0 = \sum_{i=1}^{q}g({\sf L_K} N_i,V_\alpha)\delta_{ij}
  = g({\sf L_K} N_j,V_\alpha) ~~ \forall j, \alpha $$
Similarly, evaluating (4.3) on
$\{N_j,N_l\}$ we find
\begin{eqnarray*}
0 & = & \sum_{i=1}^{q}g({\sf L_K} N_i,N_j)\delta_{il}
+ \sum_{i=1}^{q}g({\sf L_K} N_i,N_l)\delta_{ij} \\
  & = & g({\sf L_K} N_l,N_j) +
        g({\sf L_K} N_j,N_l)
\end{eqnarray*} 
Now, substituting $i=j=l$, we have $g({\sf L_K},N_j,N_i) ~~ \forall i$.  
Thus, ${\sf L_K}N_j$ is normal to $N_j$ as well as $V_{\alpha} ~~ 
\forall \alpha$, and hence, expanding ${\sf L_K}N_j$ in the basis 
$\{ V_1 , \ldots , V_p, N_1, \ldots , N_q \}$, it follows that 
$\displaystyle{{\sf L_K} N_j =\sum_{k\neq j} a^k N_k}$.   $\Box$ 
\par \noindent 
\underline{\sf Proposition 2}: 
If ${\sf K}$ is a Killing vector field on $(M,g)$, then 
${\sf L_K \omega =0}$,  
where $\omega\equiv\sigma
({\tilde V}_1\wedge\ldots\wedge{\tilde V}_p)/\mu$ - defined by the foliating
vector fields $\{V_\alpha\}$ - is the local representation of the
distribution ${\sf D}$ characterising a $p$-dimensional fluid
${\sf (D,\eta,\sigma)}$ in a spacetime 
$(M^n,g$), and $-\mu^2\equiv\det[g(V_\alpha,V_\beta)]$.
\par \noindent
\underline{Proof}: In terms of the spacelike orthonormal set $\{N_k\}$
$$\omega = \sigma (-1)^{1+pq}\star({\tilde N_1}\wedge\ldots\wedge{\tilde N_q})
\eqno (4.4)$$
where $\star$ is the Hodge operator induced by the spacetime metric $g$ and
$g(N_k,V_\alpha)=0~\forall k=1,\dots,q~~;~\forall \alpha=1,\dots,p$. Since
for any Killing field ${\sf K}$, ${\sf L_K}$ commutes with $\star$ and metric
 dual operation
\begin{eqnarray*}
{\sf L_K} \star({\tilde N_1}\wedge\ldots\wedge{\tilde N_q})
& = & \star[{\sf L_K}({\tilde N_1}\wedge\ldots\wedge{\tilde N_q})]\\
& = & \star[\sum_{j=1}^{q}(-1)^{j-1}{\widetilde {{\sf L_K} N_j}}\wedge\Pi_j]
\end{eqnarray*}
where $\Pi_j ={\tilde N_1}\wedge\ldots\wedge{\tilde N_{j-1}}
\wedge{\tilde N_{j+1}}\ldots\wedge{\tilde N_q}$.
Now, using 
$\displaystyle{{\sf L_K} N_j =\sum_{k\neq j} a^k N_k}$
from {\sf Lemma 1}, it follows that each term in the above expansion
vanishes. Hence
${\sf L_K} \star({\tilde N_1}\wedge\ldots\wedge{\tilde N_q})=0$, 
and by Lie derivation of (4.4) with respect to ${\sf K}$ 
we conclude that ${\sf L_K} \omega=0$.   $\Box$
\par
Now we derive the dynamical consequences of (4.1) with the foliating stress
tensor [(3.25)].
First, we recall that the Einstein tensor ${\sf G}$ is divergence-free,
${\nabla\cdot {\sf G}=0}$, 
and hence for any stress tensor (4.1) implies
$\nabla\cdot {\sf T}=0$.  
In our case of interest, ${\sf T}$ is the stress tensor for a
$p$-foliation, given by [(3.25)]
${\sf T}   = -\rho{(g-\sum_{k=1}^{q}{\tilde N_k}\otimes{\tilde N_k})} $.
Then, we have 
$$\nabla\cdot {\tilde {\sf T}} = - \tilde{d\rho} +
 \sum_{k=1}^{q}\{\nabla\cdot (\rho N_k)\}N_k
+ \sum_{k=1}^{q}\rho\nabla_{N_k} N_k = 0 \eqno (4.5)$$
\par \noindent
\underline{\sf Proposition 3}: For each integral manifold ($p$-brane)
of the distribution, ${\sf D}$, characterising a $p$-dimensional fluid
${\sf (D,\eta,\sigma)}$ in $(M^n,g)$, the 
mean curvature field ${\sf H} = 0$. 
\par \noindent
\underline{Proof}: The proposition involves a local assertion. Since the
distribution, ${\sf D}$, is integrable, for every point $x\in M^n$
there exists an integral manifold, ${\cal S}^p$, passing through $x$,
and (by Frobenius Theorem) there is an open neighborhood of $x$,
${\cal U}\subset M^n$, where we may choose an orthonormal moving frame
$X_1,\ldots,X_p,N_1,\ldots,N_q$ such that $\{X_a\}$ are tangent to 
${\cal S}^p$ and $\{N_j\}$ are normal to ${\cal S}^p$. Since ${\cal S}^p$
is timelike, $g(X_a,X_a)=\epsilon_a=\pm 1$. Then the mean
cuvature$^{11}$ vector field ${\sf H}$ of ${\cal S}^p\subset M^n$ 
has the following local
representation:
$${\sf H} =\sum_{j=1}^{q}\sum_{a=1}^{p}\epsilon_a
g(\nabla_{X_a}X_a,~N_j)N_j\eqno (4.6)$$ 
From (4.5), 
$g(\nabla\cdot {\tilde {\sf T}},N_j)=0$, which implies 
$$ - N_j(\rho) + \nabla \cdot (\rho N_j) + \rho \sum_{k=1}^q
g(\nabla_{N_k} N_k, N_j) = 0\eqno (4.7) $$ 
Now, inserting the following expansion
\begin{eqnarray*}
\nabla\cdot(\rho N_j) & = & N_j(\rho) + \rho\nabla\cdot N_j\\
                      & = & N_j(\rho) + \rho\sum_{a=1}^{p}\epsilon_a
g(\nabla_{X_a} N_j,~X_a) + \rho\sum_{k=1}^{q} g(\nabla_{N_k}
N_j,~N_k)\\
                      & = & N_j(\rho) - \rho\sum_{a=1}^{p}\epsilon_a
g(\nabla_{X_a} X_a,~N_j) - \rho\sum_{k=1}^{q} g(\nabla_{N_k}
N_k,~N_j)
\end{eqnarray*}
in the equation (4.7) we have 
$\sum_{a=1}^{p}\epsilon_a g(\nabla_{X_a} X_a,~N_j)=0$. Then, by (4.6),
it follows that $g({\sf H},N_j)=0$ for $j=1,\ldots,q$. Hence
${\sf H}=0$.   $\Box$    
\par\noindent
{\sf Remark}: By the above proposition, the equation of motion of a
$p$-brane, ${\cal S}$, in a multidimensional fluid ${\sf
(D,\eta,\sigma)}$ is given by ${\sf H}=0$. Also, it can be shown$^{13}$ 
that ${\sf H}$ and $\omega$ (the local representation of ${\sf D}$) are
related by
$\iota_N d\omega |_{\cal S}=-g({\sf H},~N)|_{\cal S}$,
for every vector field, $N$, normal to the $p$-brane. Then
${\sf H}=0$ implies $\iota_N d\omega |_{\cal S}=0$.    
\par \noindent
In order to derive a further dynamical consequence of (4.5), we need
the following 
\par\noindent
\underline{\sf Lemma 2}:  The world-density function, $\eta$, satisfies
$\iota_V (d\eta + \eta \lambda) = 0$, where $V$ is any vector field
tangent to the integral submanifolds of the fluid, ${\sf
(D,\eta,\sigma)}$ 
and $\lambda$ is some 1-form.
\par \noindent
\underline{Proof}:  From (4.5),  
$g(\nabla\cdot {\tilde {\sf T}},V)=0$, which implies  
\begin{eqnarray*}
V(\rho) & = & \rho\sum_{k=1}^{q} g(\nabla_{N_k} N_k,V)\\
        & = & \rho\sum_{k=1}^{q} d{\tilde N_k}(N_k,V)\\
        & = &-\rho\sum_{k=1}^{q} {\lambda^k}_k (V)
\end{eqnarray*}
where we have used 
the identity $\nabla_{N_k} {\tilde N_k} = \iota_{N_k} d{\tilde N_k}$
and the fact that the constraint 1-forms $\{{\tilde N_k}\}$ describing
the $p$-foliation must satisfy the integrability conditions [(2.2)] 
$d{\tilde N_k}=\sum_{j=1}^{q}{\lambda^j}_k\wedge{\tilde N_j}$,  
$\lambda^j_k$ being suitable 1-forms.  
Since $\rho \equiv \sigma^2\eta$, defining $\displaystyle{\lambda 
\equiv \sum_{k=1}^q \lambda^k_k} $, we have  
$[d\eta + \eta \lambda](V) = 0. ~~ \Box $ 
\par \noindent
\underline{\sf Proposition 4} : The world density function $\eta$ on 
$(M,g)$ satisfies
$d\star(\eta\omega)=0.$
\par\noindent
{\sf Proof} : From the expression for the foliation $p$-form $\omega$ [(4.4)]
\begin{eqnarray*}
d\star(\eta\omega) & = & \sigma (-1)^{1+pq}
\{\eta d({\tilde N_1}\wedge\ldots\wedge{\tilde N_q}) \\
                   &   & + d\eta\wedge({\tilde N_1}\wedge\ldots\wedge
{\tilde N_q})\}
\end{eqnarray*}
Using the integrability conditions (2.2) for $\{{\tilde N_k}\}$ we compute
\begin{eqnarray*}
d({\tilde N_1}\wedge\ldots\wedge{\tilde N_q}) & = & \sum_{k=1}^{q}
(-1)^{k-1}d{\tilde N_k}\wedge\Phi^k \\
& = & (\sum_{k=1}^{q} {\lambda^k}_k)\wedge
({\tilde N_1}\wedge\ldots\wedge{\tilde N_q})
\end{eqnarray*}
where
$\Phi^k\equiv  {\tilde N}_1\wedge\ldots\wedge
{\tilde N}_{k-1}\wedge{\tilde N}_{k+1}\ldots\wedge{\tilde N_q}$.
Then we have
$$d\star(\eta\omega) =  \sigma (-1)^{1+pq}
(d\eta + \eta\sum_{k=1}^{q} {\lambda^k}_k)
\wedge({\tilde N_1}\wedge\ldots\wedge{\tilde N_q})$$
Since the 1-form $(d\eta + \eta\sum {\lambda^k}_k)$
does not have any tangential components (by lemma 2) and its
$\{{\tilde N_k}\}$-components do not contribute in the above equation,  
$d\star(\eta\omega)=0$.   $\Box$
\par\noindent
\underline{\sf Remark}: Using the coderivative operator
$\delta = (-1)^{n(k+1)}\star d\star$ on differential $k$-forms in a
Lorentzian manifold $(M^n,g)$, we can write $d\star (\eta\omega)=0$
as $\delta(\eta\omega)=0$.
\par
Thus our program - of investigating a $p$-dimensional fluids
as a source of spacetime curvature - would be to solve
the Einstien equations (4.1) with the stress tensor ${\sf T}$ in (3.15), 
and to specify $g$ and $(\omega,\eta)$ -- the local
representation of the fluid. In the next section 
we shall work out a complete solution for a fluid characterised by 
a 2-dimensional distribution on a spacetime $(M^4,g)$. 
\vskip 0.6truecm
\leftline{\large {\sf 5 Spherically Symmetric 2-Foliation}}
\par
As an application of our results in the previous sections we consider a
2-foliation (due to string world-sheets) of a static spherically symmetric
 spacetime $(M^4,g)$ where, in the
local chart $(t,r,\theta,\phi)$, metric $g$ is of the form
$$g=-h^2(r)dt\otimes dt +f^2(r)dr\otimes dr + r^2(d\theta\otimes d\theta
+ \sin^2\theta d\phi\otimes d\phi)\eqno (5.1)$$
Using an orthonormal basis (5.1) can be written as
$$g=-e^0\otimes e^0 + e^1\otimes e^1 + e^2\otimes e^2 + e^3\otimes e^3\eqno
(5.2)$$
where coframes are
$$e^0  =  h(r)dt;~~
e^1 = f(r)dr;~~
e^2 = rd\theta;~~
e^3 = r\sin\theta d\phi\eqno (5.3)$$
with the dual basis given by
$$X_0 = (1/h)\partial_t;~~
X_1 = (1/f)\partial_r;~~
X_2 = (1/r)\partial_\theta;~~
X_3 = (1/r\sin\theta) \partial_\phi\eqno (5.4)$$
such that $e^a(X_b)={\delta^a}_b.$
It is clear from (5.1) that $(M^4,g)$ has four Killing vector fields :
\renewcommand {\theequation}{5.5}
\begin{eqnarray}
{\sf K_0} & = & \partial_t\nonumber\\
{\sf K_1} & = & \sin\phi\partial_{\theta} +
 \cot\theta\cos\phi\partial_{\phi}\nonumber\\
{\sf K_2} & = & -\cos\phi\partial_{\theta} +
 \cot\theta\sin\phi\partial_{\phi}\nonumber\\
{\sf K_3} & = & \partial_{\phi}
\end{eqnarray}
\par \noindent 
\underline{\sf Foliation 2-Form}:  If the spacetime specified by (5.1) 
is to be foliated by string world-sheets,
we must find the appropriate 2-form $\omega\in \Lambda^2(M)$ satisfying 
[Proposition 2]
$${\sf L_{K_i}}\omega=0\eqno (5.6)$$
where the Killing vector fields $\{{\sf K_i}\}$ are given in (5.5).
The most general 2-form satisfying (5.6), on the chosen spacetime[(5.1)],
must be
$$\omega = c_1(r)dt\wedge dr + c_2(r)\sin\theta d\theta\wedge d\phi \eqno
 (5.7)$$
Locally, $\omega$ is required to be decomposable, and for timelike foliation 
$\omega$ must satisfy ${\cal G}(\omega,\omega) < 0$. These two
constraints together with (5.7) uniquely (up to a scalar function) specify
the structure of the foliation 2-form so that
$$\omega = c_1(r)dt\wedge dr\eqno (5.8)$$
Without any loss of generality we can normalise (5.8) by
${\cal G}(\omega,\omega)=-1$, and the foliation 2-form is then given by
$$\omega  =  h(r)f(r)dt\wedge dr
        =  e^0\wedge e^1\eqno (5.9)$$
where we used the equations $(5.1) - (5.4)$. It is now easy to see that the
contraint form $\star\omega$ satisfies the integrability condition [(2.3)]:
$$d\star\omega  =  - d(e^2\wedge e^3)
              =  (2/r)dr\wedge\star\omega\eqno (5.10)$$
Thus $\omega$, indeed, determines a 2-foliation. Now, introducing string
rest-mass per unit length, $\sigma$, the foliation 2-form for the extended
particle flow $(\omega,\eta)$ is written as
$$\omega = \sigma e^0\wedge e^1\eqno (5.11)$$
\par \noindent 
\underline{\sf Stress Tensor}:  From (3.15) and (5.11), the associated 
stress tensor is given by
\renewcommand {\theequation}{5.12}
\begin{eqnarray}
{\sf T} & = & \eta{\cal G}(\iota_a\omega,\iota_b\omega)e^a\otimes e^b\nonumber\\
        & = & \eta\sigma^2\{{\cal G}(e^1,e^1)e^0\otimes e^0
                            + {\cal G}(-e^0,-e^0)e^1\otimes e^1\}\nonumber\\
        & = & \eta\sigma^2\{e^0\otimes e^0 - e^1\otimes e^1\}
\end{eqnarray}
where ${\sf T}$ is expanded in the orthonormal basis given in (5.3) and (5.4).
The density function, $\eta$, can be obtained from 
Proposition 4 and (5.11):
$$0  =  d\star(\eta\omega)
   =  -\sigma d(\eta r^2)\wedge\sin\theta d\theta\wedge d\phi \eqno (5.13) $$
Then (5.13) implies $\partial_r(\eta r^2)=0$ and hence $\eta r^2$ is constant.
 Thus the
density function for 2-foliation is given by
$$\eta = c/r^2\eqno (5.14)$$
where $c$ is some positive constant since $\eta$ is defined to be positive.
\par \noindent 
\underline{\sf Solution to Einstein's Equations}:  For complete specification 
of the string-field flow
we must find the functions $h(r)$ and $f(r)$ from the Einstein equation (4.1)
with the stress tensor in (5.12). For convenience (4.1) is written in the
following form :
$${\sf P}_a  =  {\sf T}_{ab}e^b - (\Gamma/2)g_{ab}e^b\eqno (5.15)$$
where ${\sf P}_a\equiv {\sf Ric}(X_a,X_b)e^b$ are the Ricci 1-forms, and
$\Gamma\equiv {\sl trace}{\sf T}= - 2\eta\sigma^2$ by (5.12). Now, computing the
Ricci forms with respect to an orthonormal basis[(5.3), (5.4)], we find
\renewcommand {\theequation}{5.16}
\begin{eqnarray}
{\sf P}_0 & = & (1/f^2)[(h''/h) - (h'/h)(f'/f) +(2/r)(h'/h)]e^0\nonumber\\
{\sf P}_1 & = & -(1/f^2)[(h''/h) - (h'/h)(f'/f) -(2/r)(f'/f)]e^1\nonumber\\
{\sf P}_2 & = & [(1/rf^2)\{-(h'/h)+(f'/f)\}+(1/r^2)\{1-(1/f^2)\}]e^2\nonumber\\
{\sf P}_3 & = & [(1/rf^2)\{-(h'/h)+(f'/f)\}+(1/r^2)\{1-(1/f^2)\}]e^3
\end{eqnarray}
From (5.2), (5.12), (5.15) and defining $\rho\equiv\eta\sigma^2$,  we also have
$${\sf P}_0 = 0;~~
{\sf P}_1 = 0;~~
{\sf P}_2 = \rho e^2;~~
{\sf P}_3 = \rho e^3\eqno (5.17)$$
Then (5.16) and (5.17) imply
$$(h''/h) - (h'/h)(f'/f) +(2/r)(h'/h)=0\eqno (5.18)$$
$$(h''/h) - (h'/h)(f'/f) -(2/r)(f'/f)=0\eqno (5.19)$$
$$(1/rf^2)\{-(h'/h)+(f'/f)\}+(1/r^2)\{1-(1/f^2)\}=\rho\eqno (5.20)$$
To solve these equations, first, we note that subtracting (5.19) from (5.18)
gives
$$(h'/h) + (f'/f) =0\eqno (5.21)$$
Integrating (5.21) and choosing the integration constant to be $0$, we have
$$hf=1\eqno (5.22)$$
Now inserting (5.21) in (5.18) and defining $a(r)\equiv h^2 (r)$ we get
$$ a'' + (2/r)a' = 0\eqno (5.23)$$
The general solution to (5.23) is found to be
$$a(r)\equiv h^2(r)=(\beta - 2m/r)\eqno (5.24)$$
where $\beta$ and $m$ are constants with $m > 0$. Then using (5.21), (5.22) and
 (5.24)
in the equation (5.20) we find
$$(1/r^2)(1-\beta)=\rho\eqno (5.25)$$
It is clear from (5.25) that
$$\beta\neq 1 \Leftrightarrow \rho\neq 0\eqno (5.26)$$
Hence for non-vanishing string field flow $(\omega,\eta)$, the constant $\beta$
can not be equal to 1. Comparing (5.25) with (5.14)
and using $\rho\equiv\eta\sigma^2$ we also have
$$\beta = 1 - c\sigma^2\eqno (5.27)$$
Collecting our results in (5.22), (5.24) and (5.27) the spacetime metric
[(5.1)] is now given by
\renewcommand {\theequation}{5.28}
\begin{eqnarray}
g & = & - (1 - c\sigma^2 - 2m/r)dt\otimes dt + (1 - c\sigma^2 - 2m/r)^{-1}
dr\otimes dr\nonumber\\
  &   & + r^2(d\theta\otimes d\theta + \sin^2\theta d\phi\otimes d\phi)
\end{eqnarray}
and from (5.22) the foliating 2-form $\omega$[(5.11)] takes the following
form :
$$\omega = \sigma dt\wedge dr\eqno (5.29)$$
Thus the equations
(5.28), (5.29) and (5.14) completely determine the local flow of a
2-dimensional fluid
generated by radial strings in a static spherically symmetric spacetime.
The metric in (5.28) may be interpreted as representing a spacetime associated
 with
a particle of mass $m$ (at $r=0$) surrounded by spherically symmetric
distribution of strings with density $\eta = c/r^2$ [(5.14)].
\vskip 0.5truecm
\leftline{{\sf Properties of the Solution (5.28) :}}
\par
(a) For $c\sigma^2 < 1$, this solution has a horizon of radius
$$r_0 = 2m/(1 - c\sigma^2)\eqno (5.30)$$
The equation (5.30) shows that the Schwarzschild radius for the mass $m$ is
enhanced by a factor $(1 - c\sigma^2)^{-1} > 1$.
\par
(b) For $m=0$, there is no horizon but the spacetime has a 
{\sf naked} singularity
at $r=0$. To see this, we substitute $p=2$ (dimension of foliation) and
$n=4$ (spacetime dimension) in (4.2) to find the Ricci (scalar) curvature :
$${\sf R} = 2c\sigma^2/r^2\eqno (5.31)$$
We also compute an invariant scalar constructed from the curvature
2-forms ${\sf R}_{ab}$, where $a,~b~=0,~1,~2,~3$ :
$$\star({\sf R}^{ab}\wedge\star {\sf R}_{ab})
=24m^2/r^6 + 8m(c\sigma^2)/r^5 + 2(c^2\sigma^4)/r^4\eqno (5.32)$$
Besides the existence of the singularity at $r=0$, 
(5.31) and (5.32) also imply that
the spacetime remains curved with $m=0$ - that is - with the foliating
strings alone, and the metric, (5.28),
remains well-behaved even with $m=0$ and $c\sigma^2=1$.
\par
(c) Using (5.2), if we rewrite the foliation stress tensor, (5.12), in
the form,
${\sf T} = -\rho (g - e^2\otimes e^2 - e^3\otimes e^3)$,
then for any causal vector $V$ we have
\renewcommand {\theequation}{5.33}
\begin{eqnarray}
{\sf T}(V,V) & = & -\rho\{g(V,V) - [g(X_2,V)]^2 - [g(X_3,V)]^2\}\nonumber\\
             & \geq & -\rho g(V,V)
              =  (1/2)({\sl trace}{\sf T})g(V,V)
\end{eqnarray}
where we have used the inequalities $\rho\geq 0$ and $g(V,V)\leq 0$.
The equation (5.33) shows that the stress tensor (5.12) satisfies the
{\sl strong energy condition}, and this also means that the gravitational field
- generated by the foliating string world-sheets - is attractive.
\par
Thus the fluid of string world-sheets gives rise to a non-trivial
solution (to Einstein's equations) - which is non-flat, static and spherically
symmetric with a naked singularity. 
Such a network of line-like objects could be used to model a
multilayer star where the constituents of each layer follows different equation
of state.
\vskip 0.5truecm
\leftline{{\sf Modelling Star with 2-Foliation :}}
\par
Following a suggestion in reference [12], we consider a two-layer star in
which the {\sf core} consists of a (spatially isotropic) perfect fluid, and
the exterior is formed by a spherically symmetric distribution of strings as in
 (5.29).
Thus the stress tensor for the core is given by
$${\sf T}_c=(\rho_c + \nu){\tilde u}\otimes{\tilde u} + \nu g_c\eqno (5.34)$$
where $u$ is a unit timelike vector field - called flow vector field,
$\rho_c$ is the density of the core and taken to be a constant, and $\nu$ is
the spatially isotropic pressure function.
Then, in the chart $(t,r,\theta,\phi)$, the metric tensor $g_c$
for the core of radius $r_c$
is given by a special case $(\rho_c\equiv constant)$ of the
well-known Oppenheimer-Volkov$^{14}$ solution to the Einstein field equations :
\renewcommand {\theequation}{5.35}
\begin{eqnarray}
g_c & = & -(1/4)[3(1-{\frac 13}{\rho_c}r_c^2)^{\frac 12}
                 -(1-{\frac 13}{\rho_c}r^2)^{\frac 12}]^2 dt\otimes
 dt\nonumber\\
    &   & +(1-{\frac 13}{\rho_c}r^2)^{-1}dr\otimes dr
        +r^2(d\theta\otimes d\theta + \sin^2\theta d\phi\otimes d\phi)
\end{eqnarray}
where $r\in (0,r_c)$ and $u=\partial_t$ in (5.34). Futhermore, the isotropic
pressure $\nu(r)$ can be obtained from the Oppenheimer-Volkov$^{14}$
equation :
$$(3\nu+\rho_c)^2/(\nu+\rho_c)^2 = [(3\nu_0+\rho_c)^2/(\nu_0+\rho_c)^2]
                                   (1 - {\frac 13}{\rho_c}r^2)\eqno (5.36)$$
where $\nu_0$ is the pressure at $r=0$.
\par
The metric for the
spacetime region foliated by the radial strings [(5.29)] is taken as
[(5.28)] :
\renewcommand {\theequation}{5.37}
\begin{eqnarray}
g_s & = & - (1 - q - 2m/r)dt\otimes dt + (1 - q - 2m/r)^{-1}
dr\otimes dr\nonumber\\
    &   & + r^2(d\theta\otimes d\theta + \sin^2\theta d\phi\otimes d\phi)
\end{eqnarray}
where $q\equiv c\sigma^2$ [from (5.28)] is a positive constant.
In (5.37) we require $r\in (r_c,r_s)$ with $r_s$ as star-radius, and $r_s>r_c$.
For $r>r_s$, we have the Schwarzschild {\sf vacuum} metric :
\renewcommand {\theequation}{5.38}
\begin{eqnarray}
g_v & = & - (1 - 2m_v /r)dt\otimes dt + (1 - 2m_v /r)^{-1}dr\otimes dr\nonumber\\
    &   & + r^2(d\theta\otimes d\theta + \sin^2\theta d\phi\otimes d\phi)
\end{eqnarray}
To complete our model we need to match these metrics continuously across the
boundaries of different layers. The matching condition for $g_s$ and $g_c$
at $r=r_c$ is given by $g_s\bigg|_{r_c}=g_c\bigg|_{r_c}$, and hence from (5.35)
and (5.37) we have
$$2m = r_c({\frac 13}{\rho_c}{r_c^2}- q)\eqno (5.39)$$
Similarly, from (5.37) and (5.38), the continuity of $g_s$ and $g_v$ at $r=r_s$
implies
$$2m_v  =  2m + q r_s
    =  {\frac 13}{\rho_c}{r_c^3} + q(r_s - r_c)\eqno (5.40)$$
where $r_s>r_c$. Now, if we impose the condition [see (5.35)]
$${\frac 13}{\rho_c}{r_c^2}<1\eqno (5.41)$$
then it follows from (5.39), (5.40) and (5.41) that $q<1$, and
$$r_c  >  2m/(1 - q)~~~;~~~
r_s  >  2m_v\eqno (5.42)$$
The above conditions, (5.41)-(5.42), ensure that the spacetime regions
specified by the corresponding metrics [(5.35), (5.37) and (5.38)] are static
and free from any singularities. We also remark that (5.41) naturally follows
from the condition that prohibits gravitational collapse of the core. To see
this, first we note that the pressure $\nu=0$ at the core-surface $r=r_c$, and
this implies [from (5.36)]
$$1 = [(3\nu_0+\rho_c)^2/(\nu_0+\rho_c)^2](1 - {\frac 13}{\rho_c}r_c^2)\eqno
 (5.43)$$
From the above equation (5.43), we find
$${\frac 13}{\rho_c}r_c^2 = 4\nu_0(2\nu_0+\rho_c)/(3\nu_0+\rho_c)^2\eqno
 (5.44)$$
The right side of (5.44) can be easily seen to be an increasing function of the
central pressure $\nu_0$. However, evaluating the limit of (5.44) as
$\nu_0\rightarrow\infty$, we get
$${\frac 13}{\rho_c}r_c^2\bigg|_{\nu_0\rightarrow\infty} = {\frac 89}\eqno
 (5.45)$$
The equation (5.45) shows that there exists a maximum
$R_m\equiv r_c\bigg|_{\nu_0\rightarrow\infty}$ for the
core-radius $r_c$ with the given density $\rho_c$, and hence
$${\frac 13}{\rho_c}r_c^2\leq{\frac 13}{\rho_c}R_m^2<1\eqno (5.46)$$
The above inequality provides the validity of the condition (5.41) which leads
to the model of a non-collapsing star with spherically symmetric distribution
of strings.
\par
Our discussion on 2-foliation
suggests that a complete general-relativistic theory of multidimensional
perfect fluids may enable us to investigate a new class of collapse
problems, the possible formation of horizons and the nature of the associated
singularities.
\par\noindent
\leftline{\large {\sf 6 Conclusion :}}
\par
We emphasize that the results in this paper strictly
follow from the concepts introduced in {\sf Definition 4} 
which offer a precise description of $p$-dimensional fluids. 
Consequently, the
foliation $p$-form $\omega$
[with ${\cal G}(\omega,\omega) < 0$] together with the {\sl uniqueness} property
 of
stress-energy tensors directly lead to our formulation of
the dynamics and symmetries of such self-gravitating systems.
Then the local decomposable $p$-form $\omega$ which defines a timelike 
$p$-foliation of a
spacetime 
and the {\sl world density} function $\eta$, enable us to introduce 
the local 
number-density [(3.1)] of
the leaves of the foliation. They also give rise to
the concept of spatial volume (and energy)
of a $p$-brane with respect to any {\sl observer}
[remarks (a), (b) and (c) after (3.7)]. These two ingredients then
naturally 
motivates a precise definition of the stress-energy tensor 
${\sf T}$[(3.15)] for 
multidimensional perfect fluids.
\par It is interesting to
observe, from the equation [(4.2)] relating {\sf trace(T)} and the
the scalar curvature ${\sf R}$, that a 2-dimensional spacetime ($n=2$) does not
admit any {\sl massive} particle ($p=1$) flow, however, only massless flows are
consistent with the field equations. In fact, a slight modification of 
our {\sf Definition 4} permits
construction of null foliations, where $\omega$ satisfies
${\cal G}(\omega,\omega)=0$.
\par Furthermore,
$\omega$ provides the local description of the dynamics 
[{\sf Proposition 3} and {\sf Proposition 4}] of $p$-branes 
and carries the symmetries [{\sf Proposition 1} and {\sf Proposition 2}] 
of a foliated 
spacetime, and we have demonstrated,
in section 5, that the spacetime symmetries
determine to a large extent the structure of foliations [(5.7)] and
hence the associated stress tensor.
This example suggests the posssibility of other
non-trivial solutions to the Einstein equations in a spacetime region dominated
by extended objects like strings and membranes. 
\par\noindent
\leftline{\large {\sf Appendix} : {\sf Definition of} ${\cal G}$ }
\par\noindent
Given a semi-Riemannian manifold $(M^n,g)$,  
the symmetric bilinear form ${\cal G}$ on 
the vector space $\Lambda^{p}(M)$ of differential $p$-forms on $M$ is, first, 
defined$^{9}$ on
decomposable $p$-forms and then the definition is extended to any $p$-forms
by linearity. Consider two decomposable $p$-forms $\Theta$ and $\Phi$ given by
$$\Theta  =  \alpha_1\wedge\ldots\wedge\alpha_p~~~;~~~
\Phi    =  \beta_1\wedge\ldots\wedge\beta_p$$
where $\alpha_i,~\beta_i\in\Lambda^{1} (M)$. 
Then, by definition,
${\cal G}(\Theta,\Phi)\equiv\det[g({\tilde \alpha_i},{\tilde \beta_j})]$ --
where ${\tilde \alpha}$ is the metric dual of the 1-form $\alpha$. Now, 
for any pair of $p$-forms $\omega$ and $\chi$, 
${\cal G}$ has the following useful properties :
$$\omega\wedge\star\chi ={\cal G}(\omega,\chi)\star 1~~~;~~~
{\cal G}(\star\omega,\star\chi)=(-1)^s {\cal G}(\omega,\chi)$$
where $\star 1$ is the volume form on $M$ induced by the metric $g$, and $s$ is
the index of $g$.
\newpage
\centerline{\large {\sf References }}
\vskip 1.0truecm
\begin{description}
\item[{$^1$}] T W B Kibble, {\sf Topology of Cosmic Domains and
Strings}, J. Phys. {\bf A9} (1976), 1387-1398.
\item[{$^2$}] H Nielsen and Olesen, {\sf Vortex-Line Models for Dual
Strings}, Nucl. Phys. {\bf B61} (1973), 45-61.
\item[{$^3$}] I Y Kobzarev, L B Okun and Y B Zel'dovich, {\sf
Cosmological Consequences of a Spontaneous Breakdown of a Discrete
Symmetry}, Sov. Phys.-JETP {\bf 40} (1975), 1-5.
\item[{$^4$}] G 't Hooft, {\sf Magnetic Monopoles in Unified Gauge-
Theories}, Nucl. Phys. {\bf B79} (1974), 276-284. \newline
              A Polyakov, {\sf Particle Spectrum in Quantum Field
Theory}, JETP. Lett. {\bf 20} (1974), 194-195.
\item[{$^5$}] C T Hill, D N Schramm and J N Fry, {\sf Cosmological
Structure Formation from Soft Topological Defects}, Comments Nucl. Part. 
Phys. {\bf 19} (1989), 25-39. \\
              M Mukherjee, {\sf Gravitational Fields of Cosmic
Membranes}, Class. Quant. Grav. {\bf 10} (1993), 131-146. 
\item[{$^6$}] R K Sachs and H Wu, {\sf General Relativity for Mathematicians}
             (Springer-Verlag, 1977). 
\item[{$^7$}] B O'Neill, {\sf Semi-Riemannian Geometry} (Academic Press, 1983).
\item[{$^8$}] M Crampin and F A E Pirani, {\sf Applicable Differential 
Geometry} 
(Cambridge University Press, 1986).
\item[{$^9$}] W Boothby, {\sf An Introduction to Differentiable Manifolds and 
Riemannian Geometry} (Academic Press, 1975).
\item[{$^{10}$}] P J E Peebles, {\sf Principles of Physical Cosmology} 
(Princeton University Press, 1993), 376-377.
        See also reference [1].
\item[{$^{11}$}] S Kobayashi and K Nomizu, {\sf Foundation of Differential 
Geometry} 
(Wiley Interscience, 1969).
\item[{$^{12}$}] J Stachel, {\sf Thickening the String I.  The String
Perfect Dust}, Phys. Rev. {\bf D21} (1980), 2171-2181.
\item[{$^{13}$}] J Simons, {\sf Minimal Varieties in Riemannian
Manifolds}, Ann. Math. {\bf 88} (1968), 62-105.
\item[{$^{14}$}] J L Martin, {\sf General Relativity} (Horwood, 1988).
\end{description}
\end{document}